# How much does context affect the accuracy of AI health advice?

Prashant Garg[1]*, Thiemo Fetzer[2,3]

[1]Imperial College London; London, UK.

[2]University of Warwick; Coventry, UK.

[3]University of Bonn; Bonn, Germany.

**Abstract**

Objectives

Large language models (LLMs) are increasingly used to provide health advice, yet evidence on how their accuracy varies across languages, topics and information sources remains limited. We assess how linguistic and contextual factors affect the accuracy of AI-based health-claim verification.

Methods

We evaluated seven widely used LLMs on two datasets: (i) 1,975 legally authorised nutrition and health claims from UK and EU regulatory registers translated into 21 languages; and (ii) 9,088 journalist-vetted public-health claims from the PUBHEALTH corpus spanning COVID-19, abortion, politics and general health, drawn from government advisories, scientific abstracts and media sources. Models classified each claim as supported or unsupported using majority voting across repeated runs. Accuracy was analysed by language, topic, source and model.

Results

Accuracy on authorised claims was highest in English and closely related European languages and declined in several widely spoken non-European languages, decreasing with syntactic distance from English. On real-world public-health claims, accuracy was substantially lower and varied systematically by topic and source. Models performed best on COVID-19 and government-attributed claims and worst on general health and scientific abstracts.

Discussion

High performance on English, canonical health claims masks substantial context-dependent gaps. Differences in training data exposure, editorial framing and topic-specific tuning likely contribute to these disparities, which are comparable in magnitude to cross-language differences.

Conclusion

LLM accuracy in health-claim verification depends strongly on language, topic and information source. English-language performance does not reliably generalise across contexts, underscoring the need for multilingual, domain-specific evaluation before deployment in public-health communication.

## Main Text

## Introduction

Seventeen percent of U.S. adults, including 25 % of those aged 18-29, now consult health chatbots each month (1). In Australia, almost 10 % of adults did so during the first half of 2024 (2). Running discharge notes through GPT-4 can cut the reading grade level from 11th to 6th and raise patient-understandability from 13 % to 81 % (3). Large language models (LLMs) generate text by estimating likely continuations of their training data rather than by checking against curated, clinically vetted knowledge bases, so any falsehoods embedded in prompts or upstream information sources can be reproduced and amplified.

During COVID-19, widely publicised misinformation made these risks concrete: U.S. President Donald Trump suggested injecting disinfectant or using ultraviolet light inside the body (4); India's health minister praised cow urine as a cancer cure (5); yoga guru Baba Ramdev sold "Coronil" as a remedy (6); Chinese state media claimed Traditional Chinese Medicine cured more than 90 % of Hubei cases (7).

In early 2020 the World Health Organization warned of an "infodemic" where rumours outpaced the virus itself (8). Misinformation undermines containment, fuels unsafe self-treatment, and erodes vaccine confidence. Social media accelerates its spread: 67 % of people under 25 get news mainly from Facebook, X, or TikTok (9). In many low- and middle-income countries these platforms fill gaps left by weak formal channels, allowing rumours to outrun corrections and even reversing gains such as measles elimination when vaccination falters

LLMs could democratise access to medical knowledge but can also magnify error, so evaluation in health settings is now an active research area. Early English-only studies show ChatGPT passing the U.S. Medical Licensing Examination (10) and Google's PaLM family setting new records on the MultiMedQA suite (11), while the domain-tuned Med-PaLM 2 reaches near-clinician performance on English vignettes (12). A recent systematic review finds that 95 % of 519 LLM-for-health evaluations between 2022 and 2024 were conducted in English and rarely assessed bias or fairness (13). Surveys of multilingual LLMs likewise report that training corpora are dominated by English (and, to a lesser extent, Chinese) text, with markedly lower coverage for many world languages (16).

Several studies now examine non-English performance. Jin et al. evaluate LLM responses to healthcare queries in English, Spanish, Chinese and Hindi and document markedly lower correctness and consistency outside English (17). Other audits have assessed ChatGPT on the Chinese National Medical Licensing Examination (18), compared English and Arabic responses for infectious-disease queries (19), and examined ChatGPT's usability for translating dermatology leaflets and patient-information leaflets for patients with non-English language preferences (14). Together these studies show that cross-language disparities exist, but they typically cover a small number of languages, focus on conversational question-answering rather than claim verification, or do not distinguish performance across information sources and topics relevant to public-health practice

Here we study when current LLMs correctly verify short health claims across languages and real-world information contexts. Using legally authorised EU/UK nutrition and health claims and 9 088 journalist-vetted public-health assertions from the PUBHEALTH corpus, we evaluate seven widely used LLMs in 21 languages on binary health-claim verification. We characterise how accuracy varies (i) across languages spanning a gradient of syntactic distance from English, and (ii) across topics (abortion, health, politics, news, COVID-19) and information sources (government advisories, scientific abstracts, mainstream and partisan media, YouTube). Rather than proposing a general-purpose benchmark, we treat these datasets as a structured audit of context-dependent accuracy for health claims that resemble those circulating in global media and policy debates

**Results**

Across the 1,975 authorised EU/UK claims, mean accuracy was high in English and closely related European languages but lower in several widely spoken non-European languages (Figure 1). Accuracy declines with syntactic distance from English, and languages such as Persian, Hindi, Chinese, Korean, Arabic and Turkish lie towards the lower end of this gradient. Model-specific slopes are modest for most systems but steeper for Mistral and Phi-4. Wilson 95 % CIs are narrow (median half-width ≈ 1.7 percentage points), for example, English 0.962–0.973, Portuguese 0.973–0.982, Persian 0.878–0.898, Korean 0.912–0.929.

On the 9,088 binary-labelled claims from the PUBHEALTH corpus, overall error rates were substantially higher than for the canonical health-claim registers, reflecting the noisier, less standardised nature of these statements. Figure 2 shows inaccuracy (1 – accuracy) by model, topic and source. Overall inaccuracy ranges from roughly 0.28 for the best-performing systems to about 0.44 for the weakest, with Phi-4 consistently the least accurate. Models are most accurate on COVID-19 claims and on snippets from Fox News and CDC.gov, and least accurate on general-health topics and Nature.com abstracts. The number of claims contributing to each topic and source cell is reported in Supplementary Table S4.

We formally assessed joint effects of model, topic and source using claim-level logistic regressions with claim fixed effects and cluster-robust standard errors. Taking Gemini-2-Flash-Lite as the reference, Gemma-3, GPT-4o-mini, Mistral and Phi-4 all perform significantly worse (all $p < 0.001$), whereas GPT-4.1 performs significantly better ($\beta \approx +0.46$, $p < 0.001$) and Llama-3 is marginally better ($p = 0.08$; Supplementary Table S1). Pairwise McNemar tests on the common claim set show that 20 of 21 model pairs differ significantly at a 5 % false-discovery-rate threshold (Supplementary Table S2). Adding Flesch–Kincaid grade level as a covariate barely changes the model coefficients, suggesting that lexical complexity has only a minor role once topic and source are controlled (Supplementary Table S3)

Informal inspection of misclassified items suggests that models struggle particularly with long, hedged sentences that compress multiple findings from a study into a single abstract-like claim, whereas short declarative sentences that quote a single source (for example, “The CDC says …”) are much more likely to be classified correctly.

**Discussion**

Large language models already answer millions of health queries, yet our audit shows that their reliability varies by language, topic and source. Strong English performance on canonical health claims conceals lower accuracy in several widely spoken non-European languages and systematic gaps across public-health topics and information sources, risking deeper information inequities and blunting efforts to counter the “infodemic” that is driving AI uptake in public health

Why do all seven models score far higher on Fox News or CDC snippets than on Nature text? Sentence complexity is too minor to explain the gap. Three stronger factors stand out. (i) Training data: large, template-driven news sites flood public web corpora, whereas pay-walled Nature prose is rare. (ii) Editorial framing: media claims are short, declarative and explicitly attributed (“The CDC says …”); Nature abstracts bundle multiple, hedged findings, making truth extraction harder. (iii) Topic tuning: COVID-19 myths prompted heavy moderation and fine-tuning, while time-sensitive political and general-health claims did not receive comparable domain-specific tuning. These factors, rather than vocabulary difficulty alone, likely drive the marked domain- and topic-specific gaps that in our data are at least as large as cross-language differences.

**Limitations**. Our gold-standard labels come from EU/UK health-claim registers and the U.S.-centric PUBHEALTH corpus, so culturally specific remedies, longer guidance and non-textbook clinical narratives lie outside scope. We rely on automatically generated translations for non-English claims and on model-based verification without additional human rating, although our round-trip similarity checks suggest that translation errors are small relative to reasoning errors. Models reflect static snapshots and may drift as training data or guardrails evolve. Binary labels hide uncertainty, and our seven-model panel omits vertical or smaller systems, so reported accuracies should be interpreted as approximate lower bounds rather than definitive certification. We release all translated claims and model outputs in the replication package to enable independent manual audits and error analyses.

**Path to application**. Our findings suggest that evaluation and deployment should be explicitly tailored to languages and information contexts rather than assuming that English-language performance generalises. A practical route is to begin with focused pilots in which health agencies and fact-checking networks co-develop test suites in high-risk languages and topics (for example, vaccine safety, infectious-disease outbreaks and politically salient health issues) using local claim taxonomies that can be updated over time. Publishing version-controlled test suites (with Docker images, API stubs and sample data under permissive licences) would allow researchers and vendors to re-run the same contextual evaluations as models evolve. Retrieval-grounded chatbots could then be deployed in tightly scoped venues (such as hospital discharge kiosks or government FAQ portals), with human oversight and live drift monitoring, before any broad release.

**Research extensions**. Priorities include real-time, language- and topic-specific accuracy-drift dashboards; lightweight multilingual medical fine-tuning that explicitly targets under-served languages; automated rationale extraction for audits; systematic equity checks across dialects and accessibility formats; and red-team tests against prompt poisoning and jailbreaks in high-stakes health settings.

**Policy and ethics**. The EU AI Act classifies multilingual health chatbots as high risk, requiring conformity assessment and training-data disclosure. Our results, which show sizeable accuracy gaps across languages, topics and sources, underscore the importance of enforcing those provisions in practice. Public funders should underwrite open multilingual benchmarks and domain-tuned open-source models, making reliable AI health advice a global public good rather than an English-only privilege. Adopting these measures can turn LLMs from potential misinformation amplifiers into more reliable partners in global health communication.

## Materials and Methods

### Processing Health Claims

We downloaded the Great Britain Nutrition and Health Claims Register and the EU Register on 25 March 2025. After trimming whitespace, we grouped rows by identical English claim text, retained the first non-missing value for each descriptive field (claim type, nutrient/food category, conditions of use, regulation notes, status, entry ID), and assigned sequential IDs (claim_id = 1…1,975) to the unique, legally authorised claims that serve as ground truth. For the present analysis we restricted to claims whose status was “Authorised” and treated these as ground-truth positive examples (true label = 1); non-authorised or pending claims were excluded.

**Language set and translation.** To mirror global web traffic (W3Techs share, 18 March 2025), we selected the top 20 content languages and added Hindi, giving 21 target languages spanning Europe, Asia and the Middle East. For every claim_id we prompted GPT-4o-mini-2024-07-18 (temperature = 0) with a JSON-schema wrapper to return: (i) the claim text, (ii) the nutrient/food term, and (iii) the stated health relationship in the target language. Medical terms lacking local equivalents were allowed to remain in English. GPT-4o-mini was used only to translate and restructure claims; the underlying English claims and their authorised status remained exactly as provided by the regulatory registers. GPT-4o-mini scores 82% on MMLU, evidencing broad linguistic competence (scores available at https://openai.com/index/gpt-4o-mini-advancing-cost-efficient-intelligence/).

**Translation validation.** To guard against mistranslation of health claims, we performed a round-trip audit long used in cross-cultural and machine-translation research: every non-English claim from the UK register was back-translated into English using the same GPT-4o-mini prompt in reverse. We encoded each source–back-translation pair with OpenAI’s text-embedding-3-large model (dimension = 3,072) and computed cosine similarity. Across 41,475 claim–language pairs the mean similarity was 0.91 with every language ≥ 0.87. Re-running the pipeline with Gemini-2-Flash-Lite as the translation model yielded an almost identical distribution (mean 0.92, all ≥ 0.89). Prior work using embedding-based paraphrase detection and round-trip translation often treats cosine similarity above about 0.8–0.85 as indicating near-semantic equivalence (20,21). In that light, our scores suggest that translation noise is small, and that most downstream errors stem from model reasoning rather than gross mistranslation. All parallel translations are released in the replication package to facilitate independent manual audits.

**Model-selection.** We benchmarked seven chat-oriented systems that account for a large share of current health-query traffic: GPT-4o-mini-2024-07-18 and GPT-4.1-2025-04-14 (OpenAI), Gemini-2.0-Flash-Lite (Google DeepMind), and four open-weight models deployed via Ollama: Phi-4 (Docker image ac896e5b8b34, quantised weights ≈ 9.1 GB), Mistral-7B-Instruct (f974a74358d6, ≈ 4.1 GB), Gemma-3-Instruct (c0494fe00251, ≈ 3.3 GB) and Llama-3-8B-Instruct (365c0bd3c000, ≈ 4.7 GB). The panel spans closed APIs versus fully open licensing, parameter counts from ≈ 7 B to ≈ 1 T, US and EU provenance, and a cost range of roughly US $0.07–30 per million tokens, providing a realistic snapshot of models available to public-health stakeholders. New “reasoning-grade” models (e.g. OpenAI o3 or GPT-5) were excluded because their token costs make large-scale evaluation prohibitive, though they could later help compare models’ reasoning with the expert justifications behind the health-claim registers.

**Model-based verification.** For every {claim, language, model} tuple we ran three chat completions (temperature = 0.7, max_tokens = 50) with a language-specific system prompt that (i) cast the model as a medical-claims expert, (ii) instructed a binary output “1” for “supported by the available evidence” and “0” for “refuted or unsupported”, and (iii) supplied local labels for the three claim components (nutrient/food, health relationship, conditions of use). We extracted the terminal “0”/“1” character from each completion, discarded malformed runs, and took the majority vote as the model verdict. The temperature of 0.7 was chosen to mirror typical user-facing chatbot settings and to average over stochastic variation rather than relying on a single deterministic sample; per-model agreement rates across the three runs are reported in the replication materials. For PUBHEALTH we used an analogous prompt that asked whether each claim was true (1) or false (0), and again took the majority vote across three runs.

**Evaluation**. For the register-based experiment we retained only claims whose status was “Authorised” (true label = 1) and computed accuracy per language by comparing each majority vote to this gold standard; the task is therefore to recognise supported claims, not to discriminate between true and false authorised versus rejected claims.

**Linguistic-distance metric**. Figure 1's x-axis uses URIEL lang2vec syntactic typology scores, which measure each language's distance from English via word-order and grammar features.

Processing PUBHEALTH Dataset

The PUBHEALTH corpus (15) contains around 11,000 U.S.-centred health claims labelled "true", "false", "unproven" or "mixture"; in the original release, there are 6,176 claims labelled true and 3 570 labelled false (15), with the remainder split between mixture and unproven labels. We kept the 9 088 entries labelled strictly "true" or "false" (82.4 %) and discarded mixture and unproven entries. Each claim was presented to the same seven LLMs with an expert instruction and run three times (temperature = 0.7, max_tokens = 50). Outputs were mapped "True"→1 and "False"→0; "Unknown" or other non-binary responses were dropped; majority vote across the three runs gave the final prediction.

**Analysis**. We calculated modal accuracy per model for: (i) All 9,088 binary claims; (ii) Topics: abortion, health, politics, news, COVID-19 (keyword-matched); and (iii) Sources: nature.com, foxnews.com, nytimes.com, youtube.com, cdc.gov (URL-parsed). Inaccuracy = 1 – accuracy, and models/categories were ordered by mean inaccuracy for Figure 2.

We report Wilson 95 % CIs for each language-level proportion. Joint effects are tested with a mixed-effects logistic model. We ran a logistic regression with claim-level fixed effects.

$$logit[Pr(correct_{im})] = \alpha_{c(i)} + \beta_m + \gamma_{t(i)} + \delta_{s(i)}$$

where $\alpha_{c(i)}$ absorbs unobserved claim difficulty, $\beta_m$ captures the average performance of model *m*, and the $\gamma_{t(i)}$ and $\delta_{s(i)}$ terms control for the five focal topics and five high-traffic domains that make up Figure 2. Standard errors are clustered by claim so that the stochasticity of individual prompts is reflected in the uncertainty. Pairwise model contrasts use McNemar's $\chi^2$ on the common claim set with Benjamini–Hochberg correction across 21 tests. Full coefficient matrices (Supplementary Table S1) and McNemar tables (Supplementary Table S2) are in the replication package.

**Lexical complexity covariate.** For every PUBHEALTH claim we calculated the Flesch-Kincaid grade level and appended it to the mixed-effects logit used for Figure 2. Complete coefficient matrices appear in Supplementary Table S3.

Full Length Prompts and Data Availability

Full LLM prompts for each step embedded in Python code are available in our replication package, along with all required data, in our Zenodo repository (DOI: 10.5281/zenodo.16490216), accessible at https://zenodo.org/records/16490216.

## Figures

**Figure 1.** Accuracy of seven LLMs on authorised health claims declines in several non-European languages. Points plot mean accuracy with 95 % Wilson CIs (y-axis) against syntactic distance

from English (x-axis). In-panel annotation lists each model's mean accuracy and fitted slope, flat for most but Mistral and Phi4, which also have the lowest accuracy.

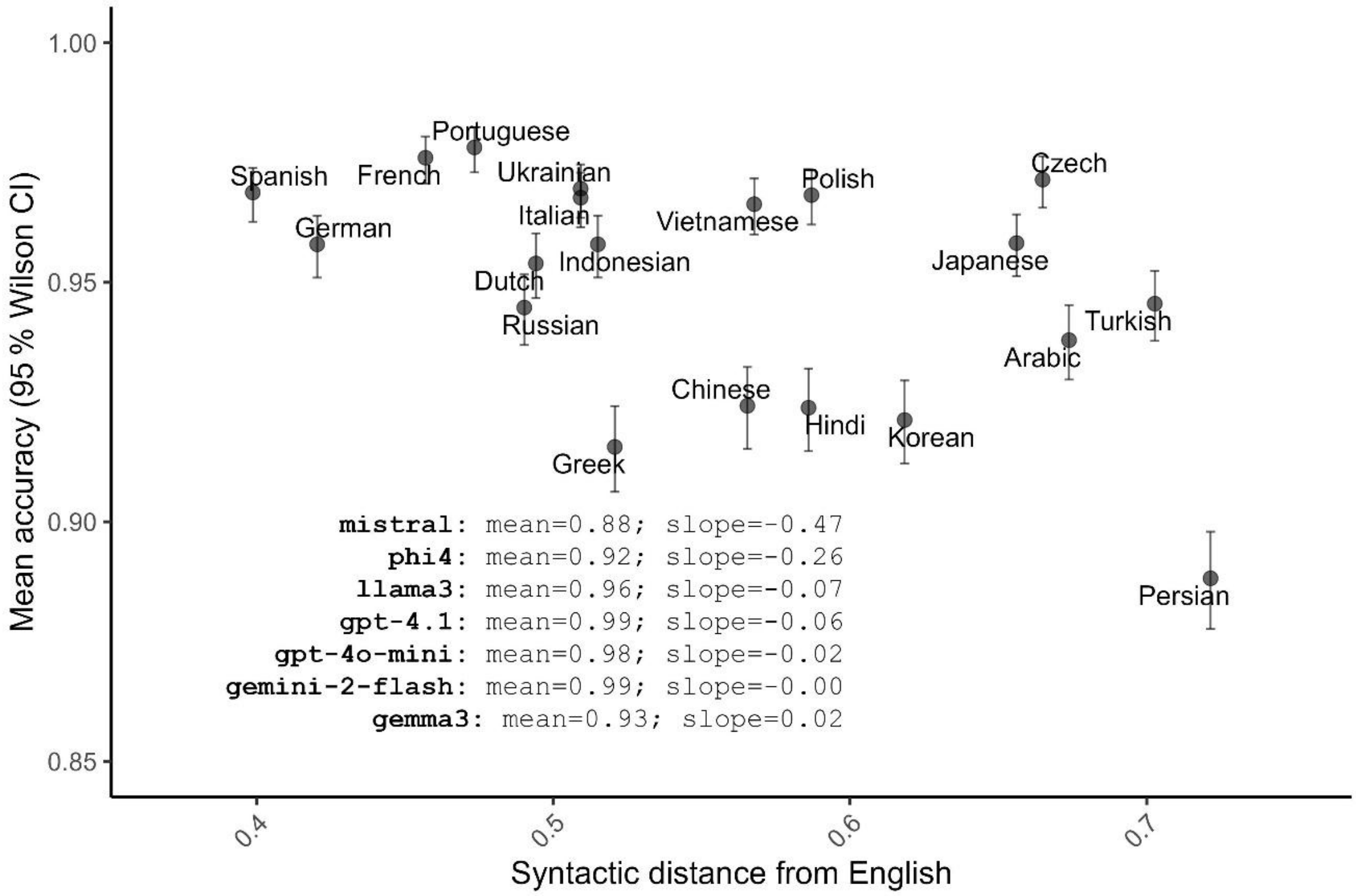

**Figure 2.** Inaccuracy heatmap (1 – accuracy; white = low, red = high) for seven LLMs on the PUBHEALTH corpus. Rows are models; columns span all 9,088 claims, five topics (abortion, health, politics, news, COVID-19) and five sources (nature.com, foxnews.com, nytimes.com,

youtube.com, cdc.gov). Models score best on Fox News and COVID-19 but falter on Nature.com and general-health items.

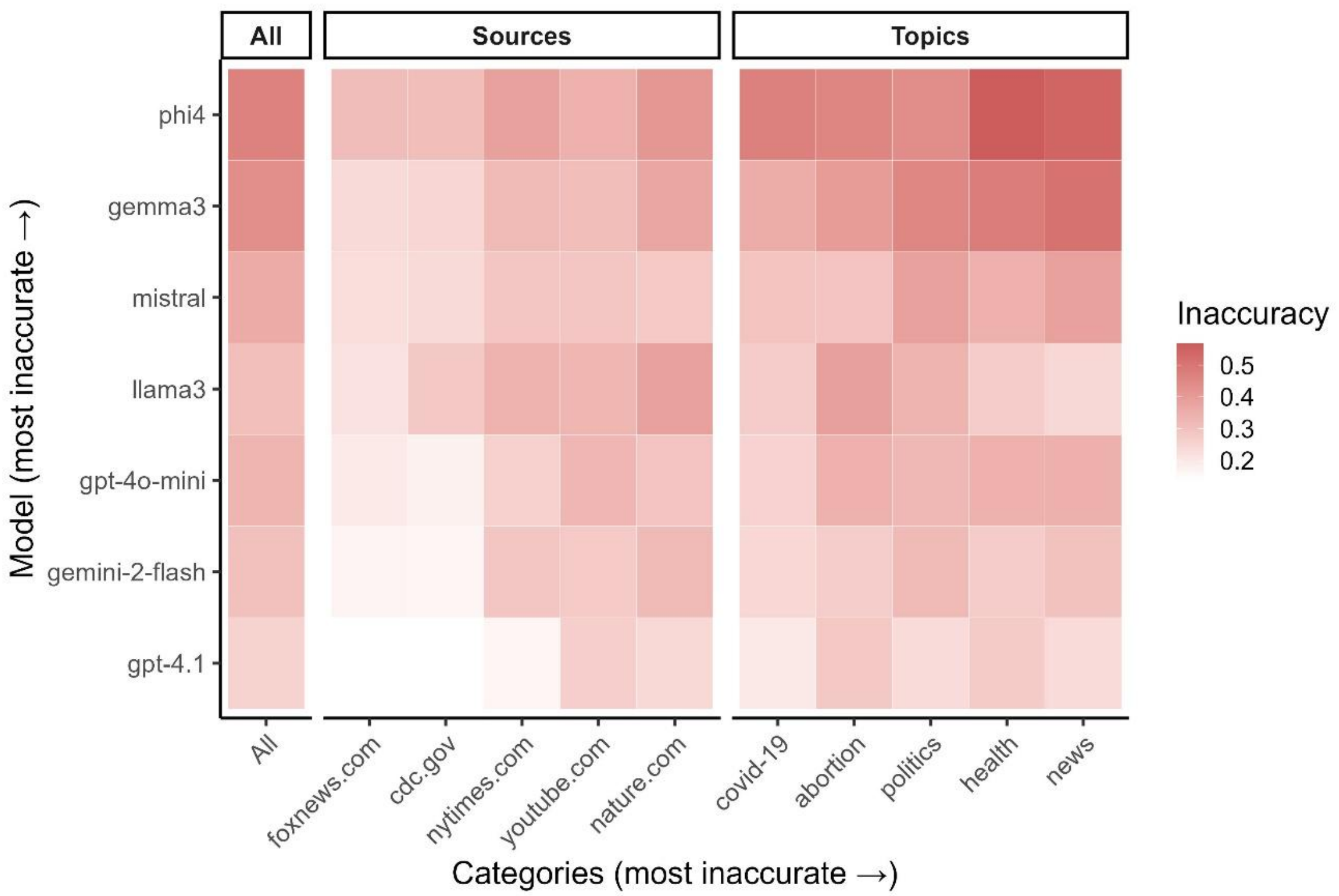

**Supplementary Information**

**Supplementary Table S1.** Coefficient table from the claim-fixed-effects logistic regression of model accuracy on the PUBHEALTH corpus. Columns list the log-odds estimate (β), cluster-robust standard error, Wald z, two-sided p, and 95% confidence bounds for each model term; Gemini-2-Flash-Lite is the reference. Claim IDs enter as fixed effects and errors are clustered by claim_id. Positive (negative) β indicates higher (lower) accuracy than the baseline. Observations = 31 851; pseudo-$R^2$ = 0.27. Reference model is Gemini-2.0-Flash.

| Term | Estimate | Std.error | Statistic | p.value | Conf.low | Conf.high |
|---|---|---|---|---|---|---|
| gemma3:latest | -1.4994 | 0.0609 | -24.6036 | 0.0000 | -1.6189 | -1.3800 |
| gpt-4.1-2025-04-14 | 0.4577 | 0.0648 | 7.0618 | 0.0000 | 0.3307 | 0.5847 |
| gpt-4o-mini-2024-07-18 | -0.3831 | 0.0573 | -6.6886 | 0.0000 | -0.4953 | -0.2708 |
| llama3:latest | 0.1315 | 0.0678 | 1.9399 | 0.0524 | -0.0014 | 0.2643 |
| mistral | -0.6486 | 0.0582 | -11.1466 | 0.0000 | -0.7627 | -0.5346 |
| phi4:latest | -1.9199 | 0.0672 | -28.5794 | 0.0000 | -2.0516 | -1.7883 |

**Supplementary Table S2.** Results of the 21 pairwise McNemar $\chi^2$ tests that compare every model duo on the identical set of PUBHEALTH claims. For each unordered pair the file reports the $\chi^2$ statistic (1 df), raw p, and Benjamini–Hochberg–adjusted p. Rows are ordered by adjusted p-value (smallest to largest). A chi-square value of $\chi^2 = 3.13$ with adj. $p = 0.077$ (final row) shows no significant difference after correction; all other 20 pairs remain significant at FDR = 5 %.

| Model A | Model B | Statistic | p value |
|---|---|---|---|
| gpt-4.1-2025-04-14 | phi4:latest | 1114.4404 | 0.0000 |
| gemma3:latest | gpt-4.1-2025-04-14 | 765.7462 | 0.0000 |
| llama3:latest | phi4:latest | 762.9986 | 0.0000 |
| gemini-2.0-flash-lite | phi4:latest | 761.9343 | 0.0000 |
| gpt-4o-mini-2024-07-18 | phi4:latest | 600.7200 | 0.0000 |
| gemini-2.0-flash-lite | gemma3:latest | 547.9317 | 0.0000 |
| gemma3:latest | llama3:latest | 515.3128 | 0.0000 |
| mistral | phi4:latest | 392.3046 | 0.0000 |
| gemma3:latest | gpt-4o-mini-2024-07-18 | 327.6049 | 0.0000 |
| gpt-4.1-2025-04-14 | mistral | 277.5307 | 0.0000 |
| gemma3:latest | mistral | 217.2552 | 0.0000 |
| gpt-4.1-2025-04-14 | gpt-4o-mini-2024-07-18 | 207.1478 | 0.0000 |
| llama3:latest | mistral | 139.6561 | 0.0000 |
| gemini-2.0-flash-lite | mistral | 120.8981 | 0.0000 |
| gpt-4o-mini-2024-07-18 | llama3:latest | 67.2116 | 0.0000 |
| gemini-2.0-flash-lite | gpt-4.1-2025-04-14 | 47.8077 | 0.0000 |
| gemma3:latest | phi4:latest | 43.9456 | 0.0000 |
| gemini-2.0-flash-lite | gpt-4o-mini-2024-07-18 | 43.5218 | 0.0000 |
| gpt-4o-mini-2024-07-18 | mistral | 22.1712 | 0.0000 |
| gpt-4.1-2025-04-14 | llama3:latest | 21.4268 | 0.0000 |
| gemini-2.0-flash-lite | llama3:latest | 3.1316 | 0.0768 |

**Supplementary Table S3.** Coefficient table from the logistic regression that adds Flesch–Kincaid grade level (fk_grade) as an additional predictor of correctness, alongside model, topic, and source, again with claim-level fixed effects and cluster-robust standard errors. Columns provide β, SE, z, two-sided p, and 95 % confidence bounds. The small positive β for fk_grade confirms that readability has only a minor influence once source and topic are controlled. Signif. codes: *** $p < 0.001$; ** $p < 0.01$; * $p < 0.05$; . $p < 0.1$.

| Term | Estimate | Std. error | Statistic | p.value | Conf.low | Conf.high |
|---|---|---|---|---|---|---|
| (Intercept) | 1.3024 | 0.2393 | 5.4433 | 0.0000 | 0.8335 | 1.7714 |
| Model: gemma3:latest | -0.7774 | 0.0316 | -24.5712 | 0.0000 | -0.8394 | -0.7154 |
| Model: gpt-4.1-2025-04-14 | 0.2412 | 0.0344 | 7.0120 | 0.0000 | 0.1738 | 0.3087 |
| Model: gpt-4o-mini-2024-07-18 | -0.2029 | 0.0305 | -6.6609 | 0.0000 | -0.2626 | -0.1432 |
| Model: llama3:latest | 0.0688 | 0.0361 | 1.9075 | 0.0565 | -0.0019 | 0.1395 |
| Model: mistral | -0.3421 | 0.0309 | -11.0752 | 0.0000 | -0.4027 | -0.2816 |
| Model: phi4:latest | -0.9853 | 0.0340 | -28.9987 | 0.0000 | -1.0519 | -0.9187 |
| Topic: covid-19 | 0.4579 | 0.1673 | 2.7367 | 0.0062 | 0.1300 | 0.7858 |
| Topic: health | -0.0678 | 0.1439 | -0.4715 | 0.6373 | -0.3499 | 0.2142 |
| Topic: news | -0.0219 | 0.1434 | -0.1526 | 0.8787 | -0.3030 | 0.2592 |
| Topic: politics | 0.2618 | 0.1595 | 1.6413 | 0.1007 | -0.0508 | 0.5744 |
| Source: foxnews.com | 0.4820 | 0.3152 | 1.5291 | 0.1262 | -0.1358 | 1.0998 |
| Source: nature.com | 2.1462 | 0.8936 | 2.4017 | 0.0163 | 0.3947 | 3.8976 |
| Source: nytimes.com | -0.1998 | 0.3948 | -0.5061 | 0.6128 | -0.9736 | 0.5740 |
| Source: Other | -0.6026 | 0.1787 | -3.3731 | 0.0007 | -0.9528 | -0.2525 |
| Source: youtube.com | -0.2618 | 0.3860 | -0.6783 | 0.4976 | -1.0183 | 0.4947 |
| fk_grade | 0.0268 | 0.0055 | 4.9110 | 0.0000 | 0.0161 | 0.0375 |